\documentclass[12pt,draftclsnofoot,onecolumn]{IEEEtran}

\usepackage{cite}
\usepackage{amssymb}
\usepackage[cmex10]{amsmath}
\usepackage{graphicx}
\usepackage{epstopdf}
\usepackage{array}
\usepackage{mathabx}
\usepackage{mathrsfs}
\usepackage{multirow}
\usepackage[caption=false,font=footnotesize]{subfig}
\usepackage{bm}
\usepackage{color}

\begin{document}

\title{Interference-Nulling Time-Reversal Beamforming for mm-Wave Massive MIMO in Multi-User Frequency-Selective Indoor Channels}

\author{Carlos~A.~Viteri-Mera and Fernando~L.~Teixeira% <-this % stops a space
\thanks{The authors are with the ElectroScience Laboratory, Department of Electrical and Computer Engineering, The Ohio State University, 1330 Kinnear Rd., Columbus, OH 
43212 USA, (614) 292-6993 (e-mail: \{viteri.5,teixeira.5\}@osu.edu).}%
\thanks{C. Viteri-Mera is also with the Department of Electronics Engineering, Univeridad de Nari\~no, Pasto, Colombia.}}%

\maketitle
\vspace{-20pt}
\begin{abstract}
Millimeter wave (mm-wave) and massive MIMO have been proposed for next generation wireless systems. However, there are many open problems for the implementation of those technologies. In particular, beamforming is necessary in mm-wave systems in order to counter high propagation losses. However, conventional beamsteering is not always appropriate in rich scattering multipath channels with frequency selective fading, such as those found in indoor environments. In this context, time-reversal (TR) is considered a promising beamforming technique for such mm-wave massive MIMO systems. In this paper, we analyze a baseband TR beamforming system for mm-wave multi-user massive MIMO. We verify that, as the number of antennas increases, TR yields good equalization and interference mitigation properties, but inter-user interference (IUI) remains a main impairment. Thus, we propose a novel technique called interference-nulling TR (INTR) to minimize IUI. We evaluate numerically the performance of INTR and compare it with conventional TR and equalized TR beamforming. We use a 60 GHz MIMO channel model with spatial correlation based on the IEEE 802.11ad SISO NLoS model. We demonstrate that INTR outperforms conventional TR with respect to average BER per user and achievable sum rate under diverse conditions, providing both diversity and multiplexing gains simultaneously.
\end{abstract}


\section{Introduction}
% Massive MIMO motivation
\IEEEPARstart{M}{assive} MIMO systems have been recently recognized as one of the technologies that can bring unprecedented performance gains for next generation wireless communications \cite{lu2014}. Among its potential benefits, noise, fading and inter-user interference (IUI) effects have been shown to progressively reduce as the number of antennas in the system increases \cite{marzetta2010}. Thus, a large number of antennas simplifies the multiple access layer and increases the system's capacity \cite{larsson2014}. However, many challenges remain for the implementation of massive MIMO systems such as: channel estimation and reciprocity issues, large pilot overheads, hardware cost, size and power limitations, network architecture adaptations, antennas and propagation aspects \cite{rusek2013}.

% mmWave massive MIMO motivation
Recently, millimeter wave (mm-wave) and massive MIMO have been proposed in tandem for next generation systems \cite{adhikary2014,swindlehurst2014}. This can be easily justified because a large number of antennas operating at mm-wave frequencies (e.g. 28, 38, 60 and 73 GHz) can be used in compact devices due to the small wavelength (4 to 10 mm approx.) and (hence) small antenna sizes. In addition, it has been shown that mm-wave networks are suitable for dense small cells (especially in indoor environments), as inter-cell interference is naturally mitigated due to high propagation losses at those frequencies \cite{boccardi2014,rappaport2013}. Another benefit of using mm-wave is the huge bandwidth availability, with some standards planning to operate with more than 2 GHz bandwidth, e.g. \cite{ieee2014}.

% Challenges in beamforming
Nevertheless, a number of problems arise for mm-wave massive MIMO systems. In particular, their performance is highly dependent on the antenna array configuration and the propagation environment. Hence, factors such as the coupling between antennas and channel spatial correlation play a significant role on the actual capacity and diversity gain that this kind of systems can achieve \cite{rusek2013}.

Beamforming is needed in mm-wave systems because of large propagation losses. Traditionally, antenna arrays use beamsteering techniques in order to increase the received power in specific directions \cite{balanis2005} and, consequently, achieve diversity gain. This is usually performed with either analog (RF) or digital (baseband) phase shifters in each antenna. Recent approaches to beamforming\footnote{The term beamforming is traditionally used to denote phased array techniques for beam steering, i.e. operating in the 2D manifold spanned by the azimuth and elevation angles. In this paper, we shall use the term beamforming in a broader sense to denote signal processing techniques that allow spatial focusing of RF power in co-range as well (3D) or even in time (4D space-time beamforming) \cite{yavuz2009,oestges2005,elsallabi2010}.} in mm-wave massive MIMO use hybrid analog beamsteering combined with digital precoding techniques assuming narrowband fading channels \cite{adhikary2014,elayach2014,alkhateeb2014,alkhateeb2014b,yang2013b}. This hybrid analog/digital solutions are necessary given that fully digital solutions requires one digital to analog converter per antenna, which is extremely constly in terms of power. However, conventional beamsteering is not always appropriate in multipath channels with frequency selective fading, such as those found in indoor environments. In those cases, more sophisticated techniques are required to take full advantage of the number of elements in the array and also multipath propagation.

The specific propagation aspects of mm-wave systems have been recently studied. Statistical models for mm-wave channels have been developed in \cite{ghosh2014,akdeniz2014}, where extremely narrow antenna radiation patterns are considered using massive MIMO. These models provide characterization of scattering clusters in the angular and delay domains, power-delay profiles (PDPs), and propagation losses in outdoor scenarios. Similar models can be found for indoor scenarios \cite{gustafson2014}. Another popular SISO channel model for indoor mm-wave systems is the IEEE 802.11ad \cite{maltsev2010}, which considers extremely narrow radiation patterns and analog beamsteering. However, there are only few studies on the spatial correlation in mm-wave MIMO channels. An interesting work is \cite{pollok2010}, where it is demonstrated that correlation at 60 GHz can be very high due to the small number of multipath components (MPCs). It has also been recognized that the specific structure of spatial correlation is highly dependent on the scattering environment. Given this conclusions, it is not clear yet whether diversity and/or multiplexing schemes should be used in order to maximize the system's gain \cite{shu2014}.

In this paper we propose a novel time-reversal (TR) based \cite{fouda2012} solution to the multi-user beamforming problem for indoor scenarios in mm-wave massive MIMO, which provides both diversity and multiplexing gains. TR is a transmission technique that enables spatial focusing of the signal at the receiver by using the time-reversed channel impulse response (CIR) as a linear filter applied to the transmitted signal \cite{oestges2005,elsallabi2010}. TR is considered a promising technique for future massive MIMO systems \cite{rusek2013}. By using TR, all multipath components are added in phase at the receiver at a specific instant providing \emph{i}) an increase in the signal power in the surroundings of the receiver (commonly referred to as spatial focusing), and \emph{ii}) a partial equalization effect (commonly known as time focusing) that reduces inter-symbol interference (ISI) caused by the channel's frequency selectivity \cite{emami2004,nguyen2006}. This features enable low computational complexity receivers, which is a key advantage of TR with respect to multicarrier (OFDM-like) systems \cite{chen2013}. Moreover, multipath components add incoherently at regions in space away from the receiver, mitigating interference to other users \cite{wang2011,viteri2014}. 

A number of works have addressed different aspects of TR beamforming, with particular focus on single user systems \cite{oestges2005,kyritsi2005,nguyen2006,wang2011,viteri2014}. In these references, the spatial and temporal focusing properties of TR have been considered, and both theoretical and empirical characterizations of bit error rate (BER) have been made under specific scenarios and channel models. A common finding in the literature is that ISI is the main limiting factor of TR. This is because ISI imposes a lower bound in the achievable BER at high signal to noise ratios (SNR) in single user systems \cite{viteri2014}.

The challenge of mitigating ISI in TR has also received increasing attention. Different equalizing solutions have been proposed in \cite{strohmer2004,nguyen2010,viteri2014} for single-user systems. An important result of \cite{viteri2014} is that the ratio between the desired signal power and the ISI power in TR increases linearly with the number of antennas. Thus, BER performance can potentially have a significant improvement when TR is applied in massive MIMO, without additional equalization.

For multiuser systems, TR for multiple access in the downlink was proposed in \cite{nguyen2005} and \cite{nguyen2006b}, where IUI is recognized as the main limiting factor of BER performance. Also, \cite{fouda2012} proposes several multi-user TR techniques. A multiple access TR technique that uses rate-backoff is proposed in \cite{han2012}, where an approximation for the signal-to-interference-plus-noise-ratio (SINR) is given, showing that it increases with the number of antennas.

However, previous works have not addressed the following aspects:
\begin{itemize}

\item Proposed beamsteering techniques in mm-wave massive MIMO are narrowband (for flat fading channels), and do not take take full advantage of multipath propagation to increase diversity gain. Thus, these techniques may not be appropriate for frequency selective channels found in indoor scenarios.%, even if the channel is sparse due to highly directive antennas.

\item TR beamforming techniques, which have been thoughtfully analyzed in other scenarios and take advantage of rich scattering, have not been studied int he context of mm-wave massive MIMO. More specifically, \cite{han2012} and \cite{viteri2014} suggest that SINR in conventional TR grows linearly with the number of antennas, enabling low complexity receivers.

\end{itemize}

In this context, the contributions of this paper are the following:

\begin{itemize}

\item We introduce a simple channel model for 60 GHz massive-MIMO, which is based on the IEEE 802.11ad model \cite{maltsev2010}. We define the probability distribution of the channel taps, their PDP, and spatial correlation.

\item We study the performance of conventional TR in multi-user systems when the number of antennas at the transmitter is very large. Moreover, we generalize the ETR \cite{viteri2014} to multi-user systems, and compare its performance with conventional TR. We demonstrate that, provided a sufficiently large number of transmit antennas, TR does not need further equalization, becoming an attractive beamforming alternative.

\item Using the previous analysis, where we find that conventional TR performance is IUI-limited, we propose a novel TR multi-user beamforming technique that minimizes IUI and exploits rich multipath commonly found in indoor environments. We call this technique interference-nulling time-reversal (INTR).

\item We analyze and compare numerically the performance of these TR techniques using the proposed statistical MIMO channel model for 60 GHz.

\end{itemize}

\subsection*{Commonly Used Acronyms in this Paper}
AP - Access Point; CB - cubicle scenario; CIR - channel impulse response; CR - conference room scenario; ETR equalized timer-reversal; INTR - interference-nulling time-reversal; ISI - inter-symbol interference; IUI - inter-user interference; LR - living room scenario; MPC - multipath component; PDP - power-delay profile; TR - time-reversal; US - uncorrelated scattering.

\subsection*{Notation}

Lower and upper case symbols represent signals in the time and frequency domains, respectively. Boldface symbols represent vectors or matrices, whose dimensions are specified explicitly. $\otimes$ is the convolution operator between two signals. $\mathbb{E}[\cdot]$ represents expectation over a random variable. The operators $(\cdot)^T$, $(\cdot)^*$, $(\cdot)^H$ and $(\cdot)^{-1}$ represent transpose, complex conjugate, Hermitian transpose, and matrix inverse, respectively. The norm of the vector $\mathbf{a}$ is denoted as $\Vert \mathbf{a} \Vert = \sqrt{\left\langle \mathbf{a}, \mathbf{a} \right\rangle} $, where $\left\langle \mathbf{a}, \mathbf{b} \right\rangle = \mathbf{b}^H \mathbf{a}$ represents the complex inner product of vectors $\mathbf{a}$ and $\mathbf{b}$. The superscripts $(\cdot)^{tr}$, $(\cdot)^{eq}$, and $(\cdot)^{in}$ denote variables calculated using time-reversal, equalized time-reversal, and interference-nulling time-reversal pre-filters, respectively.

\section{Time-Reversal Beamforming System Model}
\label{Model}

In this section, we present the general discrete signal model for TR beamforming. We first generalize to the multi-user case two TR techniques for single-user scenarios \cite{viteri2014}. These techniques serve as a baseline comparison for the novel INTR introduced in Section \ref{Model2}.% We also present the corresponding radio channel model that will be used in the next section to characterize the performance of those techniques.  

\subsection{General TR Signal Model}
\label{signalmodel}
\begin{figure}[t]
\centering
\includegraphics[width=\columnwidth]{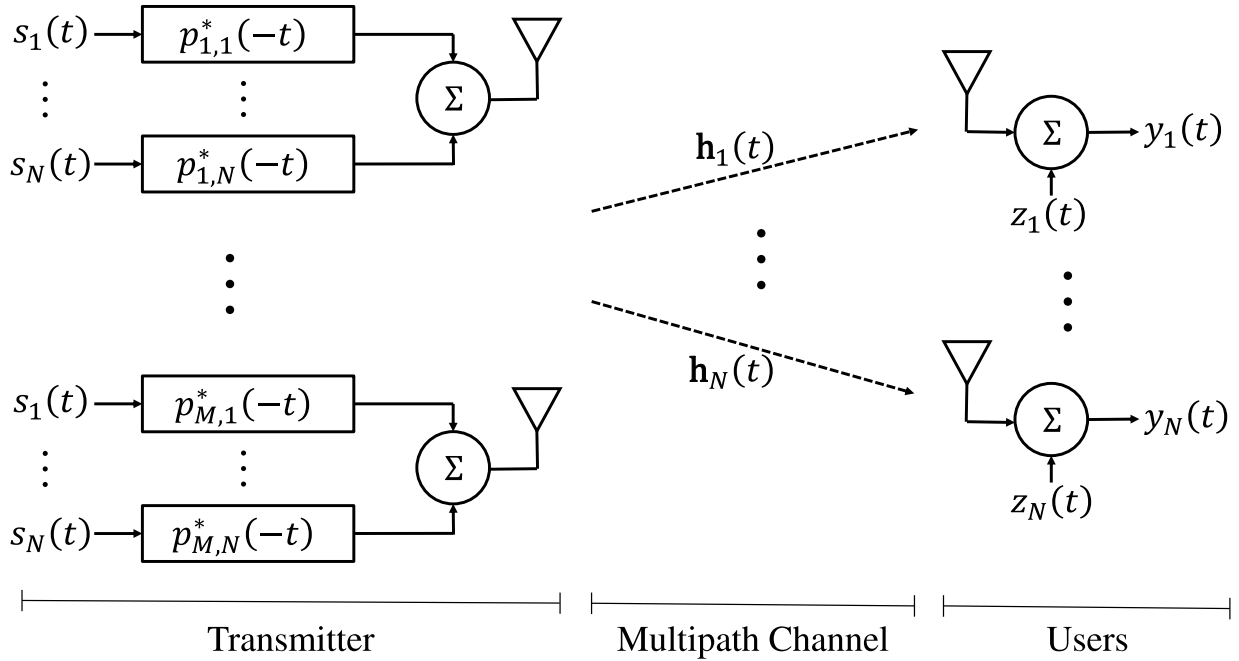}
\caption{System model. An AP with $M$ transmit antennas sends simultaneously an independent data stream to $N$ single antenna users using time-reversed pre-filters $p_{m,n}^*(-t)$.}
\label{fig_system}
\end{figure}

Consider a digital baseband downlink wireless communication system, consisting of one Access Point (AP) with $M$ transmit antennas and $N$ single-antenna user terminals as depicted in Fig. \ref{fig_system}. The transmitter has a very large number of antennas, so $M \gg N$. We denote the transmit antenna set as $\mathcal{M} = \{1,2,\ldots,M\}$ and the user set as $\mathcal{N} = \{1,2,\ldots,N\}$. Also, let $m, m' \in \mathcal{M}$ and $n, n' \in \mathcal{N}$ be arbitrary elements in those sets. The AP transmits simultaneously an independent data stream to each user. Let $s_n (t)$ be the complex random signal transmitted to the $n$-th user, where $t \in \mathbb{Z}^+$ is the discrete time index. These transmitted signals are assumed to have unit average power, i.e. $\mathbb{E}\left[ \left| s_n(t) \right|^2 \right] = 1$, $\forall n, t$, regardless of the modulation. In a TR multi-user system, the transmitter sends independent signals simultaneously to the users using different pre-filters for each one of them. Thus, the baseband transmitted signal from the $m$-th antenna is
\begin{equation}
\label{eq_transmitted}
x_m (t) = \sqrt{ \rho } \sum_{n=1}^N s_n(t) \otimes p_{m,n}^*(-t),
\end{equation}
where $\rho$ is the total average transmitted power in the AP, $p_{m,n}(t)$ is the power-normalized pre-filter from the $m$-th transmit antenna to the $n$-th user (with a duration of $L_p$ samples, i.e. $t = 0,\ldots,L_p-1$), and $h_{m,n}(t)$ is the random channel impulse response (CIR) from the $m$-th transmit antenna to the $n$-th user (with a length of $L$ samples). The random CIR vector to the $n$-th user is defined as
\begin{equation}
\mathbf{h}_n(t) = \left[h_{1,n}(t),\ldots,h_{M,n}(t)\right]^T \in \mathbb{C}^M.
\end{equation}
In Section \ref{channelmodel}, we introduce the statistical characterization of $h_{m,n}(t)$ for mm-wave channels. Let $H_{m,n}(f)$ be the discrete Fourier transform (DFT) of $h_{m,n}(t)$. In an analogous way to the time domain representation, the steering vector to the $n$-th user is
\begin{equation}
\mathbf{H}_n(f) = \left[H_{1,n}(f),\ldots,H_{M,n}(f)\right]^T  \in \mathbb{C}^M.
\end{equation}
The selection of $p_{m,n}(t)$ depends on the particular TR technique, as discussed later in this section. We define the pre-filter vector to the $n$-th user as
\begin{equation}
\mathbf{p}_n(t) = \left[p_{1,n}(t),\ldots,p_{M,n}(t)\right]^T \in \mathbb{C}^M.
\end{equation}
Let $P_{m,n}(f)$ be the DFT of $p_{m,n}(t)$. Then, we define the frequency domain pre-filter vector to the $n$-th user as 
\begin{equation}
\mathbf{P}_n(f) = \left[P_{1,n}(f),\ldots,P_{M,n}(f)\right]^T  \in \mathbb{C}^M.
\end{equation}
The received baseband signal at user $n$ is
\begin{IEEEeqnarray}{rCl}
\label{eq_received}
y_n(t) & = & \underbrace{ \sqrt{\rho} \ s_n (t) \otimes \sum_{m=1}^M p_{m,n}^*(-t) \otimes h_{m,n}(t)}_{\text{signal directed to the $n$-th user}} \nonumber \\
		  && + \underbrace{ \sqrt{\rho} \sum_{\substack{n'=1 \\ n' \neq n}}^N \sum_{m=1}^M s_{n'} (t) \otimes  p_{m,n'}^*(-t) \otimes h_{m,n}(t)}_{\text{IUI}} \nonumber \\
		  && + \underbrace{z_n(t)}_{\text{noise}},
\end{IEEEeqnarray}
where $z_n(t)$ represents additive white Gaussian noise (AWGN) with variance $\sigma_z^2$. Next, we extend the conventional TR and equalized TR single-user formulation in \cite{viteri2014} to multi-user scenarios by explicitly defining the pre-filter $p_{m,n}(t)$ in terms of the CIR. Note that $p_{m,n}(t)$ is properly normalized so the transmitted power $\rho$ regardless of the number of antennas or users.

\subsection{Multiuser Conventional TR Beamforming}
\label{trmodel}

The general idea behind TR is to use the time-reversed CIR from every antenna to the receiver as a pre-filter for the transmitted signal. Such pre-filter acts as a beamformer in the spatial domain, focusing the RF signal around the receiver. In conventional TR, assuming perfect channel state information (CSI) at the transmitter, the pre-filter vector is
\begin{equation}
\label{eq_trpilot}
\mathbf{p}^{tr}_n (t) = \frac{ \mathbf{h}_n(L-1+t)}{\sqrt{P_h^{tr}}},
\end{equation}
where $P_h^{tr}$ is a normalization factor introduced to ensure that the total transmitted power remains constant in every realization, this is
\begin{equation}
\label{eq_ph}
P_h^{tr} = \sum\limits_{n=1}^N \sum\limits_{l=0}^{L-1} \left\Vert\mathbf{h}_n(l)\right\Vert^2.
\end{equation}
Note that, in this case, the pre-filter's length is equal to the CIR length, i.e. $L_p^{tr} = L$. Replacing the conventional TR pre-filter into (\ref{eq_received}) and using the definitions in Section \ref{signalmodel}, the time domain received signal in conventional TR is
\begin{IEEEeqnarray}{lCl}
\label{eq_trreceived}
y_n^{tr}(t) \quad = \underbrace{  \sqrt{\frac{\rho}{P_h}} \sum\limits_{l=0}^{L-1} \left\Vert\mathbf{h}_n(l)\right\Vert^2 s_n (t-L+1)}_{\text{desired symbol directed to the $n$-th user}} && \nonumber \\
 + \underbrace{ \sqrt{\frac{\rho}{P_h}} \sum\limits_{\substack{ l=0 \\  l\neq L-1}}^{2L-2} \sum\limits_{m=1}^M \sum\limits_{l'=0}^{L-1} h_{m,n}(l') h_{m,n}^*(L-1-l+l') s_n (t-l)}_{\text{ISI directed to the $n$-th user}} && \nonumber \\
 + \underbrace{ \sqrt{\frac{\rho}{P_h}}\sum\limits_{\substack{ n'=1 \\  n' \neq n}}^N \sum\limits_{m=1}^M  h_{m,n'}^*(L-1-t) \otimes h_{m,n}(t) \otimes s_n' (t)}_{\text{IUI}} && \nonumber \\
 + \underbrace{ z_n(t)}_{\text{noise}}. &&
\end{IEEEeqnarray}
This received signal is composed of four terms: \emph{i}) the desired symbol multiplied by a real factor resulting from coherent combination of multipath components in the CIR, \emph{ii}) ISI caused by incoherent addition of CIR components, \emph{iii}) IUI caused by the signals directed to other users (whose TR pre-filters do not match the CIR to the $n$-th user), and \emph{iv}) AWGN. Thus, in a conventional multiuser TR beamforming system, ISI and IUI are important problems that hamper detection. In the single-user scenario, ETR was proposed before as a solution to mitigate the ISI component in the received signal \cite{viteri2014}. We extend ETR to the multiuser case next.

\subsection{Multi-user Equalized TR Beamforming}
\label{etrmodel}

ETR uses the TR pre-filter in cascade with a ZF pre-equalizer in order to mitigate the ISI of conventional TR. In \cite{viteri2014}, it is demonstrated that ETR outperforms conventional TR with respect to BER in a single-user scenario, with a marginal loss in the spatial focusing capability. We now extend this technique to the multi-user scenario by defining the pre-filter vector components for the $n$-th user as
\begin{equation}
\label{eq_etrpilot}
p^{eq}_{m,n} (t) = \frac{ h_{m,n}(L-1+t) \otimes g_n^*(-t)}{\sqrt{P_h^{eq}}},
\end{equation}
where $g_n(t)$ represents a ZF linear equalizer with length $L_E$. Thus, we have $L_p^{eq} = L+L_E-1$. The normalization factor is
\begin{equation}
\label{eq_pg}
P_h^{eq} = \sum\limits_{m=1}^M \sum\limits_{n=1}^N \sum\limits_{l=0}^{L+L_E-2} \left| h_{m,n}^*(L-1-l) \otimes g_n(l) \right|^2.
\end{equation}
One equalizer is required for each user, with the $n$-th equalizer designed to satisfy
\begin{equation}
\label{eq_zf}
g_n(t) \otimes \sum_{m=1}^M h_{m,n}^*(L-1-t) \otimes h_{m,n}(t) = \delta (t-t_0),
\end{equation}
where $t_0$ is an arbitrary delay. Equation (\ref{eq_zf}) can be written as an over-determined system of linear equations on $g_n(t)$, $t = 0,\ldots,L_E-1$. Thus, perfect ZF equalization is not possible with a finite equalizer's length \cite{hayes1996}, but a good approximation can be achieved with a sufficiently large $L_E$, eliminating the second term in (\ref{eq_trreceived}). A detailed discussion on this subject is provided in \cite{viteri2014}. Using the ETR pilot, and assuming perfect equalization, the time domain received signal at user $n$ is
\begin{IEEEeqnarray}{lCl}
\label{eq_etrreceived}
 y_n^{eq}(t) \quad = \underbrace{ \sqrt{\frac{\rho}{P_h^{eq}}} \ s_n (t-t_0)}_{\text{signal directed to the $n$-th user}} && \nonumber \\
 + \underbrace{ \sqrt{\frac{\rho}{P_h^{eq}}} \sum\limits_{\substack{ n'=1 \\  n' \neq n}}^N \sum\limits_{m=1}^M s_{n'} (t) \otimes  g_{n'}(t) \otimes h_{m,n'}^*(L-1-t) \otimes h_{m,n}(t)}_{\text{IUI}} && \nonumber \\
 + \underbrace{ z_n(t)}_{\text{noise}},&&
\end{IEEEeqnarray}
which has no ISI term, but still contains IUI. Thus, both conventional TR and ETR performance is limited by ISI and/or IUI, as detailed next.

\subsection{Performance Analysis of TR and ETR}
\label{analysis}
We now turn our attention to the power components in (\ref{eq_trreceived}) and (\ref{eq_etrreceived}), following the same procedure as in \cite{viteri2014}. The fundamental assumptions are that the system operates under uncorrelated scattering with uncorrelated channels between users, and normalized channel power, as stated in Section \ref{channelmodel}. In the derivations below, we also employ the approximation $\mathbb{E}[a/b]\approx\mathbb{E}[a]/\mathbb{E}[b]$ for two random variables $a$ and $b$, as analyzed in \cite{han2012,viteri2014} for TR systems. Complete derivations are not shown due to space constraints. Here, we verify that TR is a suitable technique for massive MIMO systems, where $M \gg 1$. Let $P_s^{tr}$, $P_{isi}^{tr}$, and $P_{iui}^{tr}$ represent the power in the first, second, and third terms in (\ref{eq_trreceived}), respectively. Then, the average desired signal power is
\begin{IEEEeqnarray}{lCl}
\label{eq_trps}
\mathbb{E} \left[ P_s^{tr} \right] =  \mathbb{E} \left[ \frac{\rho}{P_h^{tr}} \left| \sum\limits_{l=0}^{L-1} \left\Vert\mathbf{h}_n(l)\right\Vert^2 s_n (t-L+1) \right|^2 \right] \approx  \frac{M\rho\Gamma}{N}. &&\nonumber\\
\end{IEEEeqnarray}
The average ISI power in (\ref{eq_trreceived}) can be approximated as
\begin{IEEEeqnarray}{rCl}
\label{eq_trpisi}
\mathbb{E} \left[ P_{isi}^{tr} \right] &  = &  \mathbb{E} \left[ \frac{\rho}{P_h^{tr}} \left| \sum\limits_{\substack{ l=0 \\  l\neq L-1}}^{2L-2} \sum\limits_{m=1}^M \sum\limits_{l'=0}^{L-1} h_{m,n}  (l')  h_{m,n}^*(L-1-l+l') s_n (t-l) \right|^2 \right] \nonumber \\
 &  \approx &  \frac{\rho}{MN\Gamma} \sum\limits_{\substack{ l=0 \\  l\neq L-1}}^{2L-2} \sum\limits_{m=1}^M \sum\limits_{m'=1}^M \sum\limits_{l'=0}^{L-1} \mathbb{E} \left[ h_{m,n}(l')h_{m',n}^*(l') | \right] \times \nonumber \\
 &&  \qquad \mathbb{E} \left[ h_{m,n}^*(L-1-l+l')h_{m',n}(L-1-l+l') \right].
\end{IEEEeqnarray}
An approximation to the average IUI power is
\begin{IEEEeqnarray}{lCl}
\label{eq_trpiui}
 \mathbb{E} \left[ P_{iui}^{tr} \right] &  = &  \mathbb{E} \left[ \frac{\rho}{P_h^{tr}} \left| \sum\limits_{\substack{ n'=1 \\  n' \neq n}}^N \sum\limits_{m=1}^M  h_{m,n'}^*(L-1-t) \otimes  h_{m,n}(t) \otimes s_n' (t) \right|^2 \right] \nonumber \\
&  \approx &  \frac{\rho}{MN\Gamma} \sum\limits_{\substack{ n'=1 \\  n' \neq n}}^N \sum\limits_{l=0}^{2L-2} \sum\limits_{m=1}^M \sum\limits_{m'=1}^M \sum\limits_{l'=0}^{L-1} \mathbb{E} \left[ h_{m,n}(l')h_{m',n}^*(l')  \right] \nonumber \\
&&  \quad \times \mathbb{E} \left[ h_{m,n'}^*(L-1-l+l')h_{m',n'}(L-1-l+l') \right].
\end{IEEEeqnarray}
For ETR, the average desired signal power is bounded by
\begin{IEEEeqnarray}{lCl}
\label{eq_eqps}
\mathbb{E} \left[ P_s^{eq} \right]  =  \mathbb{E} \left[ \frac{\rho}{P_h^{eq}} \right] \leq  \frac{M\rho\Gamma}{N}. &&
\end{IEEEeqnarray}
The average IUI power in ETR has a similar form to (\ref{eq_trpiui}), but it is not shown here since it is not the focus of this work. However, we analyze it numerically in Section \ref{results}. From (\ref{eq_trps})-(\ref{eq_trpiui}), we can make the following remarks with respect to TR beamforming in massive MIMO systems:
\begin{itemize}
\item Both ISI and IUI powers are highly dependent on the propagation conditions. More specifically, power delay profiles and correlation between antennas are present in the terms of the form $\mathbb{E} [ h_{m,n}(l)h_{m',n}^*(l)  ]$. Thus, increasing spatial correlation would increase both ISI and IUI, degrading performance.
\item In the case of uncorrelated antennas, $\mathbb{E} [ h_{m,n}(l)h_{m',n}^*(l)  ]=0$ if $ m\neq m'$. Hence, the sums would only depend on the power delay profile (which is the same for all antennas), and both ISI and IUI powers would be independent of $M$.
\item Desired signal power increases linearly with $M$. Thus, in uncorrelated channels $\mathbb{E} [P_s^{tr}/P_{isi}^{tr} ] \rightarrow  \infty$ and $\mathbb{E} [P_s^{tr}/P_{iui}^{tr}] \rightarrow \infty$ as $M \rightarrow \infty$. This implies that, with a sufficiently large number of antennas, a conventional TR beamforming system is noise limited instead of interference limited.
\item However, if channels are spatially correlated (as in realistic scenarios), equalization and interference mitigation provided by TR reduce.
\item Note that, given that CIR statistics for users $n$ and $n'$ are the same, i.e. $\mathbb{E} \left[ h_{m,n}^*(l)h_{m',n}(l) \right]=\mathbb{E} \left[ h_{m,n'}^*(l)h_{m',n'}(l) \right]$ $\forall l$, then $P_{iui}^{tr}$ is larger than $P_{isi}^{tr}$ by a factor on the order of the number of users. Thus, IUI mitigation should be given priority over equalization when proposing improvements over conventional TR.
\end{itemize}

Given this characteristics of TR beamforming in massive MIMO, we now propose a novel TR extension to overcome the problems of IUI, even under highly correlated channels.

\section{Interference-Nulling Time-Reversal Beamforming}
\label{Model2}

\begin{figure}
\centering
\includegraphics[width=0.8\columnwidth]{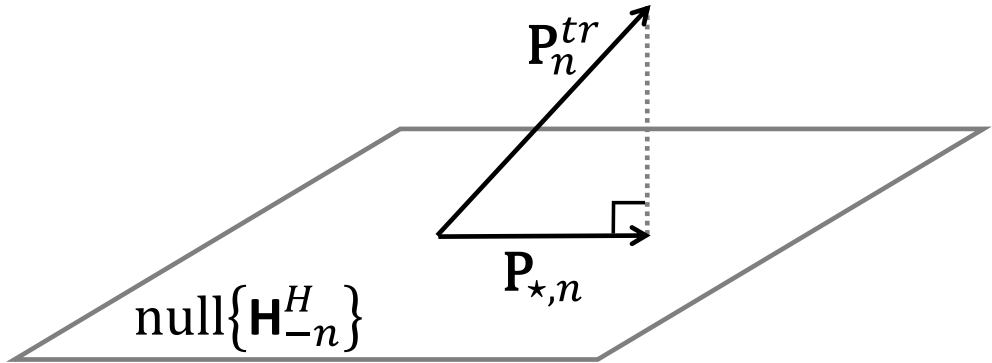}
\caption{Geometrical interpretation of the optimization procedure. The optimum pilot in the frequency domain is the conventional TR prefilter projection onto the nullspace of $\bm{\mathsf{H}}_{-n}^H$. This ensures that IUI is set to zero for every user and every frequency.}
\label{fig_geom}
\end{figure}

We are now concerned with the design of pre-filter vectors that combine the spatial focusing properties of conventional TR, while also providing additional IUI mitigation. We start from the frequency representation of the received signal, and formulate an optimization problem for the design of the pre-filters. The frequency domain equivalent of (\ref{eq_received}) is
\begin{IEEEeqnarray}{rCl}
\label{eq_receivedf}
Y_n(f) & = & \underbrace{ \sqrt{\rho} \ \left\langle\mathbf{H}_n(f),\mathbf{P}_n(f)\right\rangle S_n(f)}_{\text{signal directed to the $n$-th user}} \nonumber \\
 		  && + \underbrace{ \sqrt{\rho} \sum_{\substack{n'=1 \\ n' \neq n}}^N \left\langle\mathbf{H}_n(f),\mathbf{P}_{n'}(f)\right\rangle S_{n'}(f)}_{\text{IUI}} \nonumber \\
 		  && + \underbrace{Z_n(f)}_{\text{noise}},
\end{IEEEeqnarray}
where $S_n(f)$ is the DFT of $s_n(t)$, and $Z_n(f)$ is the DFT of $z_n(t)$. Appropriate zero padding is used in the time domain in order to represent linear convolution as a product in the frequency domain. The complex inner product defined above allows a convenient simplification in (\ref{eq_receivedf}) with respect to (\ref{eq_received}), which is useful for the problem formulation. Let $\bm{\mathsf{H}} (f) = [\mathbf{H}_1(f) \ldots \mathbf{H}_N(f)] \in \mathbb{C}^{M \times N}$ be the matrix with columns given by the steering vectors to all users. Also, let $\bm{\mathsf{H}}_{-n}(f) \in \mathbb{C}^{M \times N-1}$ be the matrix formed by removing the $n$-th column from $\bm{\mathsf{H}}(f)$, i.e. removing the steering vector to user $n$. For notational simplicity, we drop the frequency dependence in the remainder of this section. Note that the IUI power in (\ref{eq_receivedf}) is proportional to $ \sum_{n' \neq n} | \left\langle\mathbf{H}_n,\mathbf{P}_{n'}\right\rangle |^2$. Thus, our objective is to find the pre-filter $\mathbf{P}_{\star,n}$ which is closest to the conventional TR solution in the frequency domain (providing partial equalization of the received signal), and such that the IUI is set to zero. Formally, this optimization problem can be formulated as
\begin{IEEEeqnarray}{rll}
\label{eq_optim}
& & \mathbf{P}_{\star,n} = \arg \min_{\mathbf{P}_n} \quad \left\Vert \mathbf{P}_n^{tr} - \mathbf{P}_n \right\Vert^2  \nonumber \\
& & \text{subject to}  \quad  \bm{\mathsf{H}}_{-n}^H \mathbf{P}_n  = 0,  \nonumber \\
& & \forall n \in \mathcal{N}, \quad \forall f \in [0,\ldots,L+L_p-1],
\end{IEEEeqnarray}
%\begin{IEEEeqnarray}{rll}
%\label{eq_optim}
%& & \mathbf{P}_{\star,n} = \arg \min_{\mathbf{P}_n} \quad \left\Vert \mathbf{P}_n^{tr} - \mathbf{P}_n \right\Vert^2  \nonumber \\
%& & \text{subject to}  \quad \left\Vert \bm{\mathsf{H}}_{-n}^H \mathbf{P}_n \right\Vert^2 = 0,  \nonumber \\
%& & \forall n \in \mathcal{N}, \quad \forall f \in [0,\ldots,L+L_p-1],
%\end{IEEEeqnarray}
whose solution is
\begin{IEEEeqnarray}{rCl}
\label{eq_rtrpilot2}
\mathbf{P}_{\star,n} & = &  \left( \mathbf{I}_M - \bm{\mathsf{H}}_{-n} \left( \bm{\mathsf{H}}_{-n}^H \bm{\mathsf{H}}_{-n}  \right)^{-1} \bm{\mathsf{H}}_{-n}^H  \right) \mathbf{P}_n^{tr}, \nonumber \\
&& \quad \quad \quad \quad \quad \forall n, f.
\end{IEEEeqnarray}
%\begin{IEEEeqnarray}{rCl}
%\label{eq_rtrpilot}
%\mathbf{P}_{\star,n} & = &  \left( \mathbf{I}_M + \nu \bm{\mathsf{H}}_{-n} \bm{\mathsf{H}}_{-n}^H \right)^{-1} \mathbf{P}_n^{tr}, \quad \forall n, f
%\end{IEEEeqnarray}
where $\mathbf{I}_M$ is the $M \times M$ identity matrix. Thus, we call $\mathbf{P}_{\star,n}$ the interference-nulling time-reversal (INTR) pre-filter in the frequency domain. Geometrically, the constraint in the problem ensures that the vector $\mathbf{P}_{\star,n} \in \mathrm{null} \left\{ \bm{\mathsf{H}}_{-n}^H \right\}$, and the solution is the projection of $\mathbf{P}_n^{tr}$ into that null space. This is illustrated in Fig. \ref{fig_geom}. %For the selection of a suitable regularization factor, we note that the optimum prefilter satisfies $(\mathbf{P}_n^{tr} - \mathbf{P}_{\star,n}) \perp \mathrm{null} \left\{ \bm{\mathsf{H}}_{-n}^H \right\}$. Using this condition and the matrix inversion lemma, it can be demonstrated that the optimum is reached when $\nu \rightarrow \infty$ \cite[Ch. 2]{bertsekas1999}, so that (\ref{eq_rtrpilot}) reduces to
%The regularization parameter $\nu$ can be found using, for example, a line search. Also, using the matrix inversion lemma, the RTR pilot can be written as
%\begin{IEEEeqnarray}{rCl}
%\label{eq_rtrpilot2}
%\mathbf{P}_n^{re} & = &   \nu \bm{\mathsf{H}}_{-n} \left( \mathbf{I}_{N-1} + \nu \bm{\mathsf{H}}_{-n}^H \bm{\mathsf{H}}_{-n} \right)^{-1} \bm{\mathsf{H}}_{-n}^H \mathbf{P}_n^{tr}, \quad \forall n, f, \nonumber \\
%\end{IEEEeqnarray}
%which reduces the computational complexity of this solution.

\section{Channel Model for 60 GHz Massive MIMO}
\label{channelmodel}

As  mentioned above, TR actually benefits from rich scattering, so it can be conveniently applied for indoor wireless communications. In this section, we briefly describe the IEEE 802.11ad model for 60 GHz SISO systems in such scenarios \cite{maltsev2010}, and extend it to the correlated multi-user massive MIMO case. In the following, we use a statistical description of $h_{m,n}(t)$, given by its probability distribution, power delay profile (PDP), and spatial correlation in the context of massive MIMO systems.

\subsection{Channel Tap Distribution}

The most popular channel model for mm-wave propagation is the IEEE 802.11ad. This is a SISO double directional statistical channel model based on a limited set of measurements,  complemented with ray-tracing simulations. This model is defined for three indoor scenarios: conference room (CR), living room (LR), and cubicle environment (CB). Some important model features include: support of two types of antennas (isotropic and basic steerable antenna array), support of polarization, wideband and pathloss modeling under LoS and NLoS situations.

The IEEE 802.11ad channel model follows a scattering cluster structure, both in time and angular domains. Thus, several multipath components (MPCs) observed in the CIR have similar propagation delays and angles of departure/arrival. More specifically, each central ray arriving at the receiver has pre-cursor rays (which arrive earlier) and post-cursor rays (which arrive later). This is due to irregular scattering objects and geometrical features which are large compared to the wavelength. Both pre-cursor and post-cursor rays have less amplitude than the central ray. The resolvability of those MPCs depend exclusively on the system's sampling time (bandwidth). When those MPCs are not resolvable, they contribute to the same tap in the CIR. Given this propagation characteristics, we assume that $h_{m,n}(t)$ has zero mean and that $|h_{m,n}(t)|$ is Nakagami distributed, with parameters $\mathfrak{m}$ and $\Omega$, for all $m$, $n$ and $t$ \cite{dersch1993}. Recall that the $\mathfrak{m}$ parameter in the Nakagami distribution is analogous to the $K$ factor in the Rician distribution, and that a larger $\mathfrak{m}$ implies a large power ratio between the central ray (specular component) and the other rays (diffuse components). The parameter $\Omega$ depends also on the amplitudes of the specular and diffuse components, and on the channel PDP (tap average power) \cite{dersch1993}. Table \ref{tablescenarios} shows the values of $\mathfrak{m}$ and RMS delay spread in the IEEE 802.11ad scenarios. The larger value of $\mathfrak{m}$ in the CB scenario is due to the reduced scattering within the cubicles, which reduces the number and power of diffuse components contributing to each channel tap.
\begin{table}
\caption{Nakagami $\mathfrak{m}$ parameter and RMS delay spread of IEEE 802.11ad scenarios}%
\label{tablescenarios}
\centering
\begin{tabular}{ccc}
\hline \hline
Scenario 	& Nakagami $\mathfrak{m}$ parameter 	& RMS delay spread [ns]	\\
\hline
CB			& 4.34						& 3.47	\\
CR			& 2.56						& 4.82	\\
LR			& 1.74						& 7.81	\\
\hline \hline
\end{tabular}
\end{table}
\subsection{Power Delay Profile}
\begin{figure}[!t]
\centering
\includegraphics[width=0.7\columnwidth]{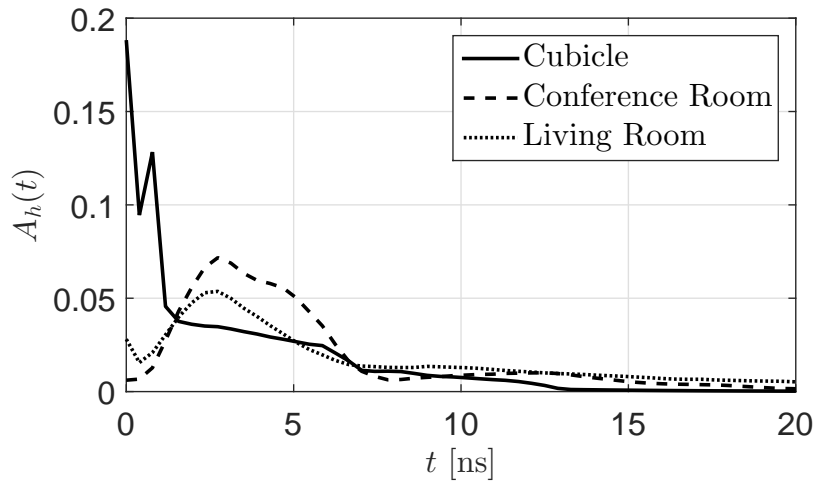}
\caption{Power delay profile of IEEE 802.11ad channel model scenarios with isotropic antennas. RMS delay spreads are 3.47 ns for the CB scenario, 4.82 ns for CR and 7.81 ns for LR.}
\label{fig_pdps}
\end{figure}

We are particularly interested in the PDP, a second order statistic defined as
\begin{equation}
\label{eq_pdp}
A_h(t) = \mathbb{E} \left[ |h_{m,n}(t)|^2 \right], \quad \forall m,n,
\end{equation}
where the expectation is calculated over CIRs that are subject to the same large-scale fading \cite{goldsmith2005}. Signal power components depend on the PDP and spatial correlation, as seen in Section \ref{analysis}. We assume that all CIR in the system have the same PDP. This is valid for mm-wave indoor environments, where APs are usually positioned on or close to the ceiling and similar shadowing affects all elements in the transmit array. We also define the following constraint on the CIR total power:
\begin{equation}
\label{eq_channelnormalization}
\sum_{t=0}^{L-1} \mathbb{E}\left[ \left| h_{m,n}(t) \right|^2 \right] = \sum_{t=0}^{L-1} A_h(t) = \Gamma,
\end{equation}
where $\Gamma \ll 1$ is a constant accounting for channel induced propagation losses. This constraint implies that all channels between the transmit antennas and each receiver have same average power. Fig. \ref{fig_pdps} shows PDPs obtained over $10^6$ realizations of the IEEE 802.11ad model for the three scenarios simulated under NLoS and isotropic antennas. We do not consider LoS situations since they correspond to flat-fading channels, which are of no interest here. Isotropic antennas  are assumed so the system can take advantage of all MPCs in the channel. In practice, planar omni-directional antennas (e.g. \cite{ranvier2008}) would be a good alternative for implementation. We observe that RMS delay spread is minimum for the CB scenario, where an AP is located in the ceiling of an office populated with cubicles. In that case, scattering is confined within the cubicle's structure and other delayed paths (e.g. reflections from outer walls) are obstructed. On the other hand, CR and LR scenarios correspond to more open spaces, where first and second order reflections from walls are considered. Those reflections cause long tails in their PDPs, increasing their delay spread.

\subsection{Spatial Correlation Model}

Consider $m, m' \in \mathcal{M}$, $n, n' \in \mathcal{N}$, and $t, t' \in \{0,\ldots,L-1 \}$.  We make the following assumptions with respect to CIRs in the systems:
\begin{itemize}

\item CIR are correlated across transmit antennas, i.e. the spatial channel autocorrelation function is $R_h(\Delta d) \neq 0$, where $\Delta d$ is the distance between two measured CIRs. The specific correlation structure depends on the array configuration, but it is assumed that the process in wide sense stationary with respect to the space. This implies that $\mathbb{E}[h_{m,n}(t)h_{m',n}^*(t)] \neq 0$.

\item Different users have uncorrelated CIRs to the AP, i.e. $h_{m,n}(t)$ and $h_{m,n'}(t)$ are uncorrelated if $n \neq n'$, $\forall m, t$. This is due to the fact that MPCs are  independent for different users. This can be clearly seen in the CB environment, where each user is assumed to be in its own cubicle.

\item CIR taps are uncorrelated, i.e. $h_{m,n}(t)$ and $h_{m,n}(t')$ are uncorrelated if $t \neq t'$, $\forall m, n$. This is the conventional uncorrelated scattering (US) assumption widely used in the literature \cite{proakis2008}, and implies that contributions to different taps come from different scatterers.

\end{itemize}

\begin{figure}[!t]
\centering
\includegraphics[width=0.8\columnwidth]{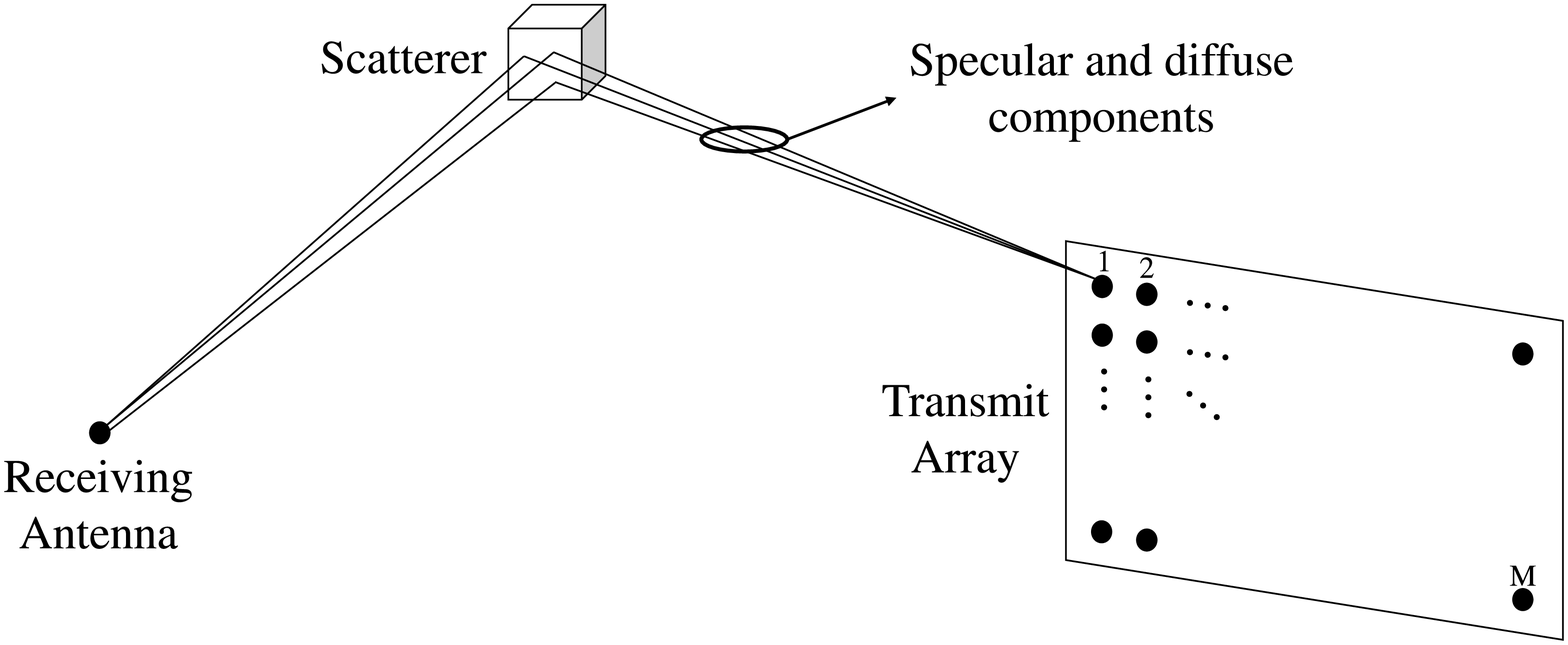}
\caption{Method to generate correlated Nakagami CIR. Different taps are assumed to have contributions from specular and diffuse reflections from different objects. The transmit array is planar (rectangular) with uniformly distributed elements.}
\label{fig_method}
\end{figure}

\begin{figure}[!t]
\centering
\includegraphics[width=0.7\columnwidth]{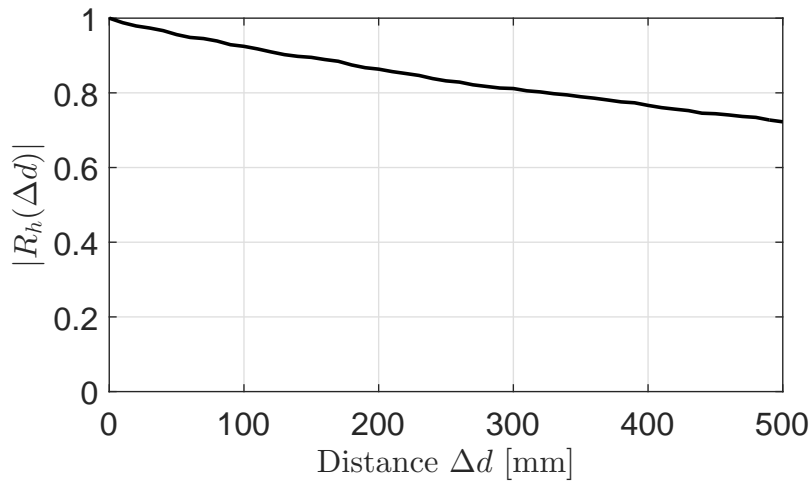}
\caption{CIR spatial correlation as a function of distance $\Delta d$, calculated over 1000 CIR realizations. }
\label{fig_correl}
\end{figure}

Nakagami correlated variables (across antennas) are generated according to the method described in \cite{dersch1993}, as follows. Consider the setting in Fig. \ref{fig_method}. A planar randomly-oriented array with $M$ isotropic elements is located in the environment according to the standard \cite{maltsev2010}. An isotropic receiving antenna is randomly located in the environment as well. Each tap is assumed to have specular and diffuse contributions from an irregular scatterer (located according to the corresponding delay), whose amplitudes depend on the PDP and the desired $\mathfrak{m}$ parameter. All contributions to a fixed tap in a given CIR come from the same scatterer, with different taps corresponding to different scatterers.  Using this procedure, the resulting normalized spatial correlation function $R_h(\Delta d)$ is shown in Fig. \ref{fig_correl}. These results are consistent with measured and simulated spatial correlations in 60 GHz channels, e.g. \cite{pollok2010}. High correlation values are caused by the reduced number of dominant MPC contributing to each tap. The specific correlation between transmit antenna elements depends only on the geometry of the array. For the numerical validation shown in Section \ref{results}, we use rectangular arrays with 32 ($8\times4$), 64 ($8\times8$), or 128 ($16\times8$) elements with a uniform separation of 20 mm.

\section{Numerical Results and Discussion}
\label{results}
In this section, we present numerical results for the performance analysis of the multiuser TR, ETR and INTR techniques, as described in Sections \ref{Model} and \ref{Model2}. 

\subsection{Pilot Length and Channel Correlation}
First, we analyze the impact of pre-filter's length $L_p$ and spatial correlation on the signal power components. We calculate the values of $P_s$, $P_{isi}$, and $P_{iui}$ for the three techniques over 1000 channel realizations with and without spatial correlation. Results are shown in Table \ref{TableChannelModel} for 2 and 10 users in the CB scenario with 64 antennas. For conventional TR beamforming, it is clear that IUI power is the main problem for multi-user communications, as it can be up to an order of magnitude greater than ISI power. It is also observed that channel correlation decreases desired signal power and increases interference, affecting the overall system performance. Also, note that when using spatially uncorrelated channels, both ISI and IUI suffer small or no change when increasing the number of antennas. However, when the CIRs are correlated both types of interference suffer a small increase. For ETR, increasing pre-filter's length improves ISI suppression, but IUI remains the same as in conventional TR. ISI reduction in ETR by increasing $L_p$ is a typical consequence of zero-forcing equalization \cite{viteri2014}. However, ETR is not designed to mitigate IUI. Thus, BER performance of TR and ETR are expected to be very similar since the scenarios we consider are clearly IUI limited.

For INTR, IUI mitigation improves by increasing $L_p$. This is due to the discarding of $L-1$ time samples when performing the transformation between the frequency domain prefilter (of length $L+L_p-1$) and the time domain prefilter (of length $L_p$). Such discarding is necessary due to the circular convolution theorem. Thus, the time domain prefilter is a least squares projection of the optimum frequency domain solution. The error in the projection reduces as $L_p$ increases.

\begin{figure*}[!t]
\centering
\subfloat[]{\includegraphics[width=0.5\columnwidth]{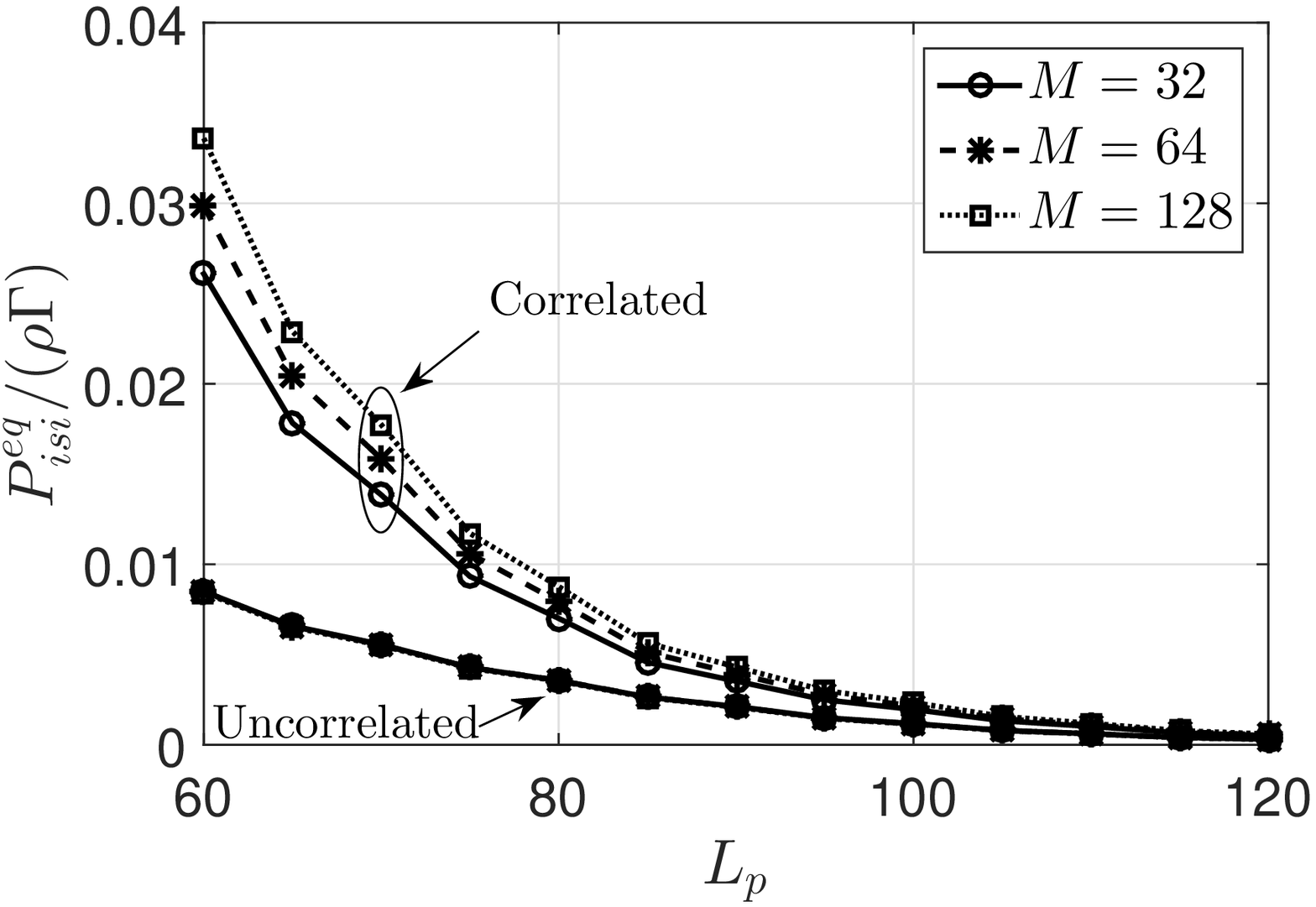}
\label{fig_isi}}
\subfloat[]{\includegraphics[width=0.5\columnwidth]{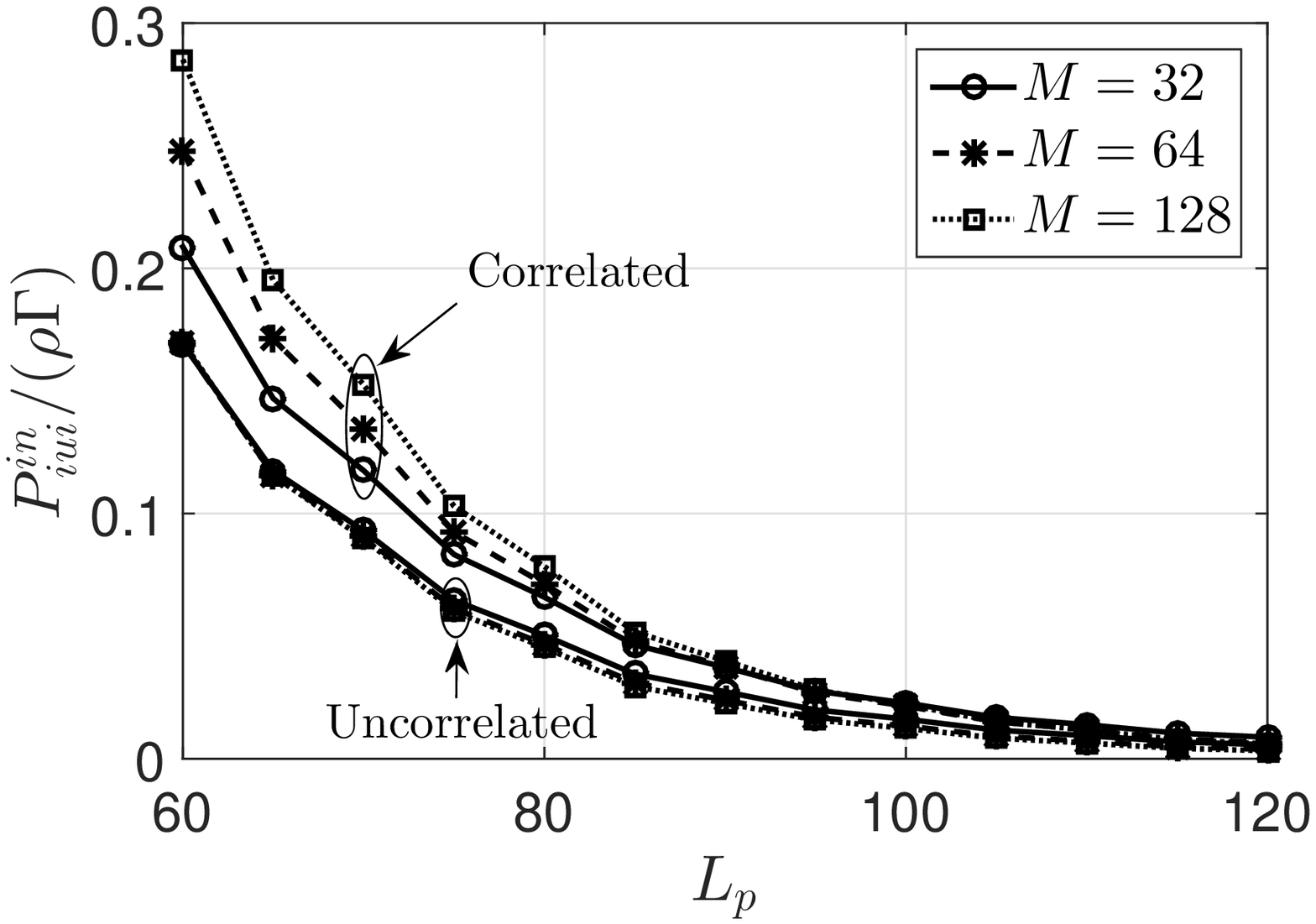}
\label{fig_iui}}
\newline
\caption{Prefilter length ($L_p$) vs (a) ISI power in ETR, and (b) IUI power in INTR. These results were obtained with $L=60$ and $N=10$ in the CB scenario. Other signal components in each technique remained approximately constant vs. $L_p$. It is noted that increasing $L_p$ reduces ISI power in ETR and IUI in INTR. This is due to the discarding of $L-1$ time samples when performing the transformation between the frequency domain prefilter (of length $L+L_p-1$) and the time domain prefilter (of length $L_p$). Such discarding is necessary due to the circular convolution theorem. Thus, the time domain prefilter is a least squares projection of the optimum frequency domain solution. The error in the projection reduces as $L_p$ increases.}
\label{fig_interference}
\end{figure*}

We observe the impact of signal power components over the BER performance in Fig. \ref{fig_interference}, where the influence of channel spatial correlation is also shown. Signal to noise ratio is defined as $\text{SNR}=\rho\Gamma/\sigma_z^2$, where $\sigma_z^2$ is the variance of $z_n(t)$ $\forall n, t$. These results were obtained for 5 users and 32 antennas in the CB scenario, with a transmission of $10^6$ BPSK symbols over 1000 channel realizations. Performance of both TR and ETR is limited by IUI, which causes a lower bound in the BER. We notice that ETR does not provide a significant improvement over conventional TR in the case of multi-user massive MIMO systems. Thus, ETR does not offer any advantage for such scenarios, given its greater computational complexity with respect to TR. In the case of INTR, IUI is successfully mitigated and hence INTR outperforms the other techniques. We also observe that channel correlation degrades system performance in all cases.

\begin{table*}[t]
\caption{Received signal power components for $M = 64$. Values are normalized $/(\rho\Gamma)$.}%
\label{TableChannelModel}
\centering
\begin{tabular}{c|c|ccc|ccc|ccc}
\hline \hline
\multicolumn{1}{c|}{\multirow{2}{*}{\centering{Number of users $N$}}} & \multicolumn{1}{c|}{\multirow{2}{*}{\centering{Technique}}} & \multicolumn{3}{c|}{$L_p = L = 60$} & \multicolumn{3}{c|}{$L_p = 90$} & \multicolumn{3}{c}{$L_p = 120$} \\
\cline{3-11}
& & $P_s$  & $P_{isi}$ & $P_{iui}$ & $P_s$  & $P_{isi}$ & $P_{iui}$ & $P_s$ & $P_{isi}$ & $P_{iui}$ \\
\hline
\multicolumn{11}{c}{Uncorrelated channel}\\
\hline
\multicolumn{1}{c|}{\multirow{3}{*}{\centering{2}}} & TR & 32 & 0.15 & 0.51 & - & - & - & - & - & - \\
& ETR & - & - & - & 31.9 & 0.01 & 0.52 & 31.9 & 0.001 & 0.52 \\
& INTR & 31.7 & 0.15 & 0.09 & 31.6 & 0.16 & 0.01 & 31.6 & 0.15 & 0.002 \\
\hline
\multicolumn{1}{c|}{\multirow{3}{*}{\centering{10}}} & TR & 6.4 & 0.03 & 0.9 & - & - & - & - & - & - \\
& ETR & - & - & - & 6.37 & 0.002 & 0.9 & 6.37 & 0.0003 & 0.9 \\
& INTR & 5.77 & 0.04 & 0.17 & 5.58 & 0.04 & 0.02 & 5.55 & 0.04 & 0.004 \\
\hline
\multicolumn{11}{c}{Correlated channel}\\
\hline
\multicolumn{1}{c|}{\multirow{3}{*}{\centering{2}}} & TR & 31.4 & 0.6 & 0.88 & - & - & - & - & - & - \\
& ETR & - & - & - & 30.3 & 0.02 & 0.88 & 30.3 & 0.003 & 0.88 \\
& INTR & 30.8 & 0.61 & 0.15 & 30.6 & 0.6 & 0.02 & 30.6 & 0.6 & 0.003 \\
\hline
\multicolumn{1}{c|}{\multirow{3}{*}{\centering{10}}} & TR & 6.39 & 0.15 & 1.48 & - & - & - & - & - & - \\
& ETR & - & - & - & 6.07 & 0.004 & 1.47 & 6.07 & 0.0005 & 1.47 \\
& INTR & 5.32 & 0.14 & 0.25 & 5.06 & 0.14 & 0.04 & 5.02 & 0.14 & 0.006 \\
\hline \hline
\end{tabular}
\end{table*}

\begin{figure}[!t]
\centering
\includegraphics[width=\columnwidth]{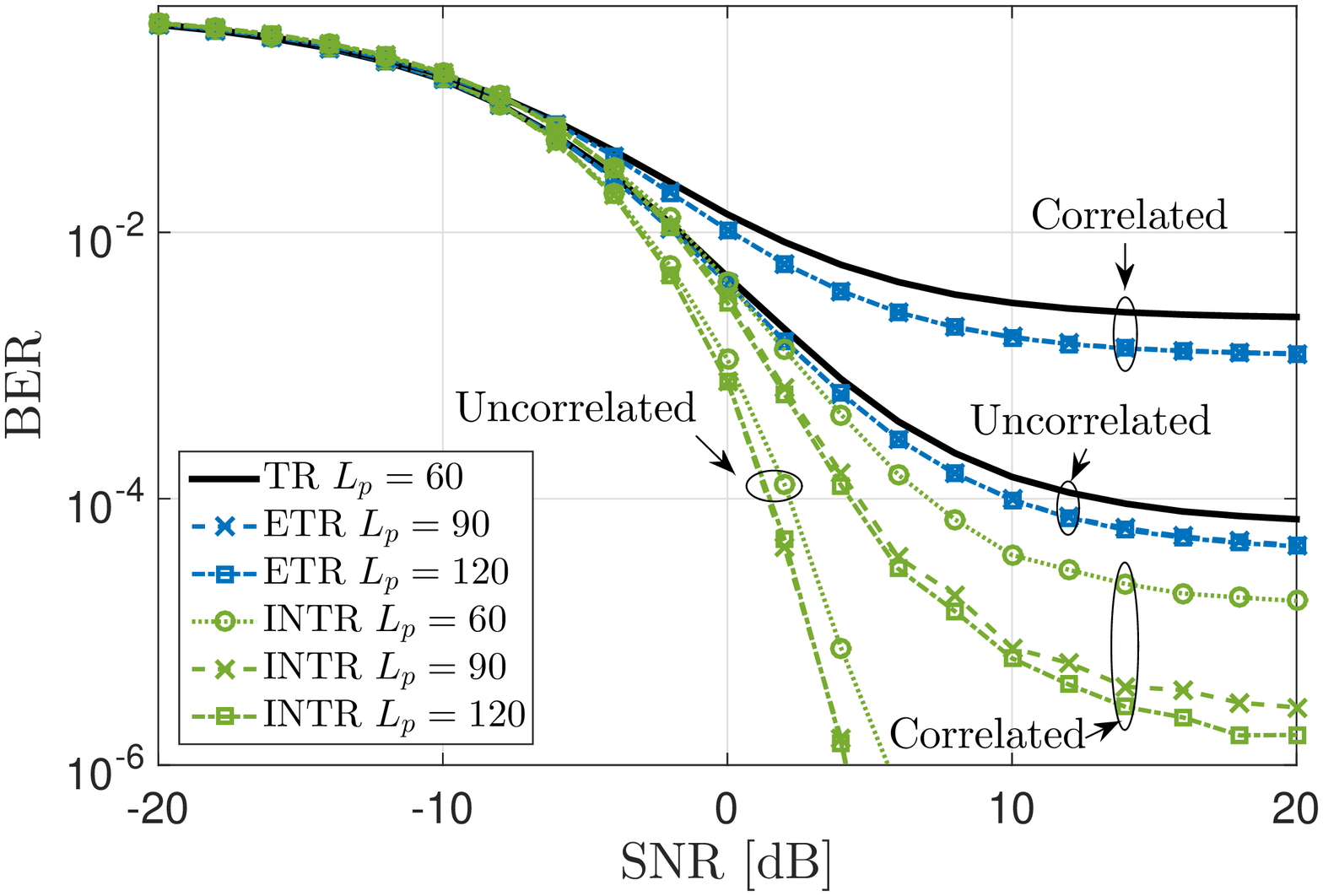}
\caption{Average BER per user comparison of TR, ETR, and INTR, under correlated and uncorrelated channels (across antenna elements) with $M=32$, and $N=5$. Results are shown for different pilot lengths ($L_p$). It is observed that spatial correlation increases ISI and IUI, degrading performance. Also, increasing prefilter's length improves IUI mitigation in INTR.}
\label{fig_correlation}
\end{figure}

\subsection{Number of Antennas and Number of Users}

\begin{figure*}[!t]
\centering
\subfloat[]{\includegraphics[width=0.5\columnwidth]{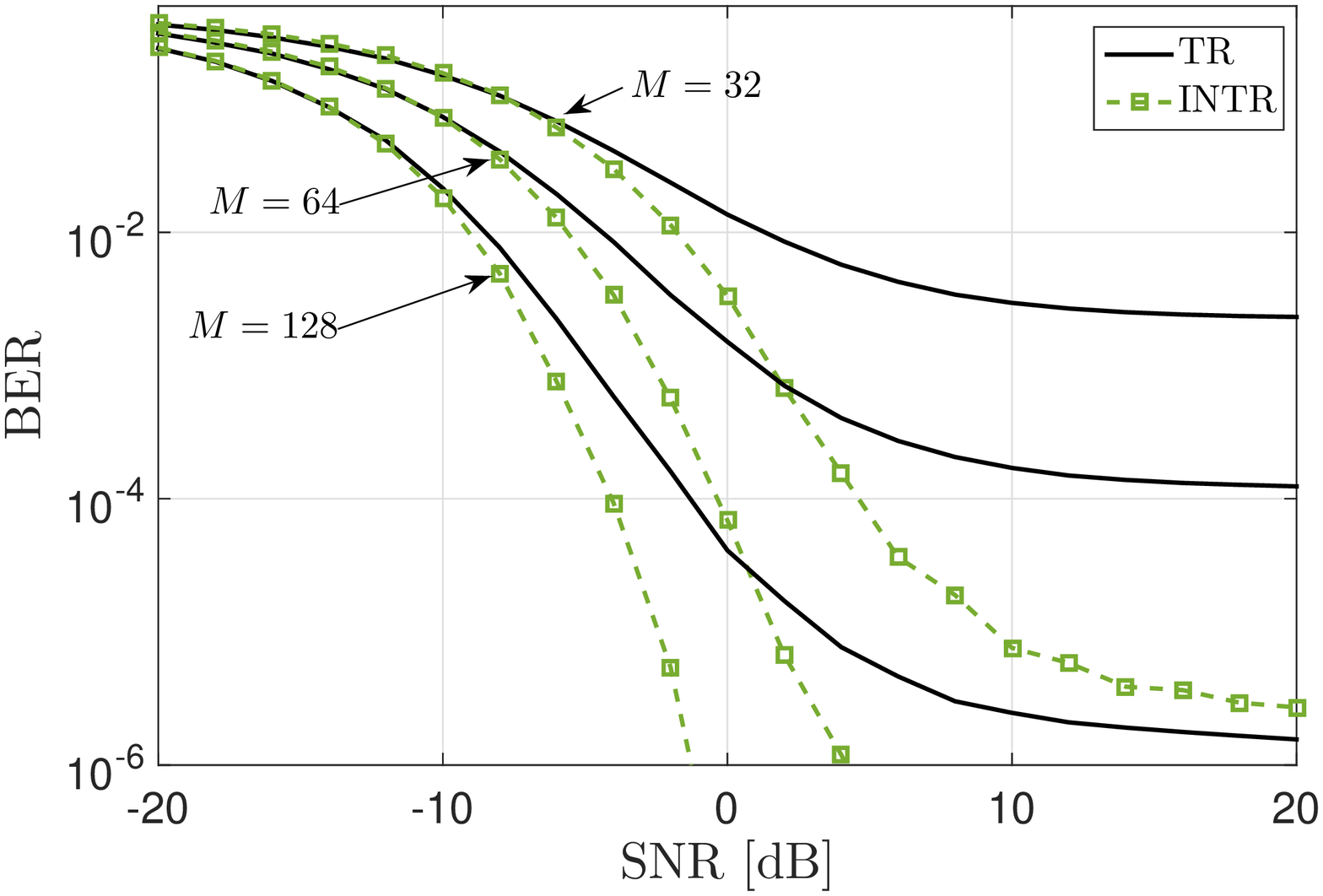}
\label{fig_antennas}}
\subfloat[]{\includegraphics[width=0.5\columnwidth]{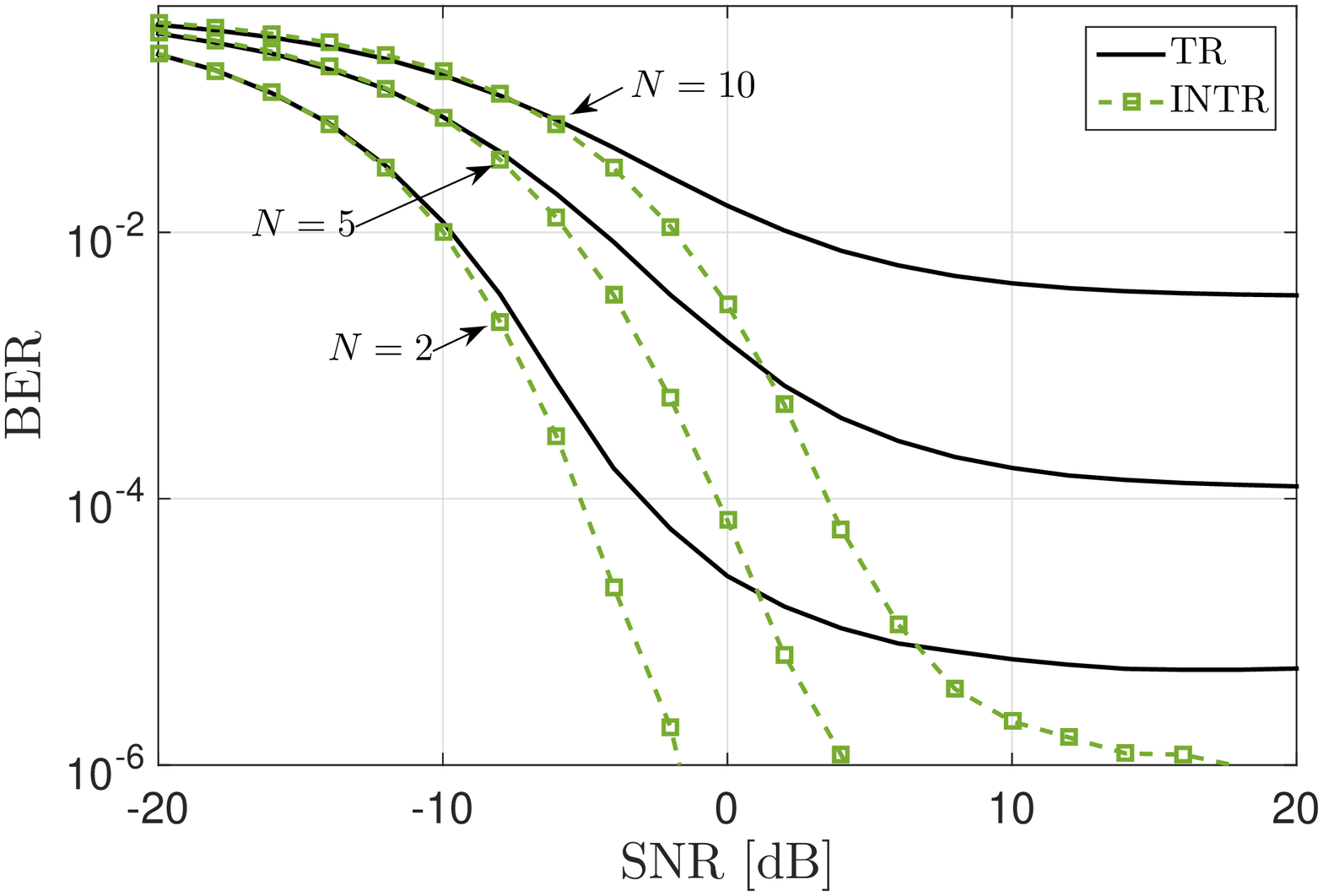}
\label{fig_users}}
\newline
\caption{Avera BER per user for TR and INTR. (a) Different number of antennas $M$ with $L_p=90$ and $N=5$. (b) Different number of users $N$ with $L_p=90$ and $M=64$. An important diversity gain is achieved even in spatially correlated channels. The effect of IUI is mitigated by increasing the number of antennas.}
\label{fig_ant-users}
\end{figure*}

Fig. \ref{fig_ant-users} shows the average BER performance results per user for varying number of antennas and users. These results where obtained with the transmission of $10^6$ BPSK symbols over 1000 spatially correlated channel realizations. We used a fixed pre-filter length $L_p =90$ and the CB scenario PDP. Conventional TR performance results are consistent with the analysis made in Section \ref{analysis}. The desired signal power increases linearly with the number of antennas while interference components remain constant. Thus, the minimum achievable BER per user improves by increasing the number of antennas, providing diversity gain. On the other hand, increasing the number of users with a fixed number of antennas decreases the desired signal power and increases IUI. This is reflected in a higher BER for larger $N$. INTR outperforms conventional TR in every simulated scenario. Nevertheless, the performance improvement provided by INTR is more evident with a large number of users or a limited number of antennas.
\subsection{Average Achievable Sum Rate}
\begin{figure*}[!t]
\centering
\subfloat[]{\includegraphics[width=0.33\columnwidth]{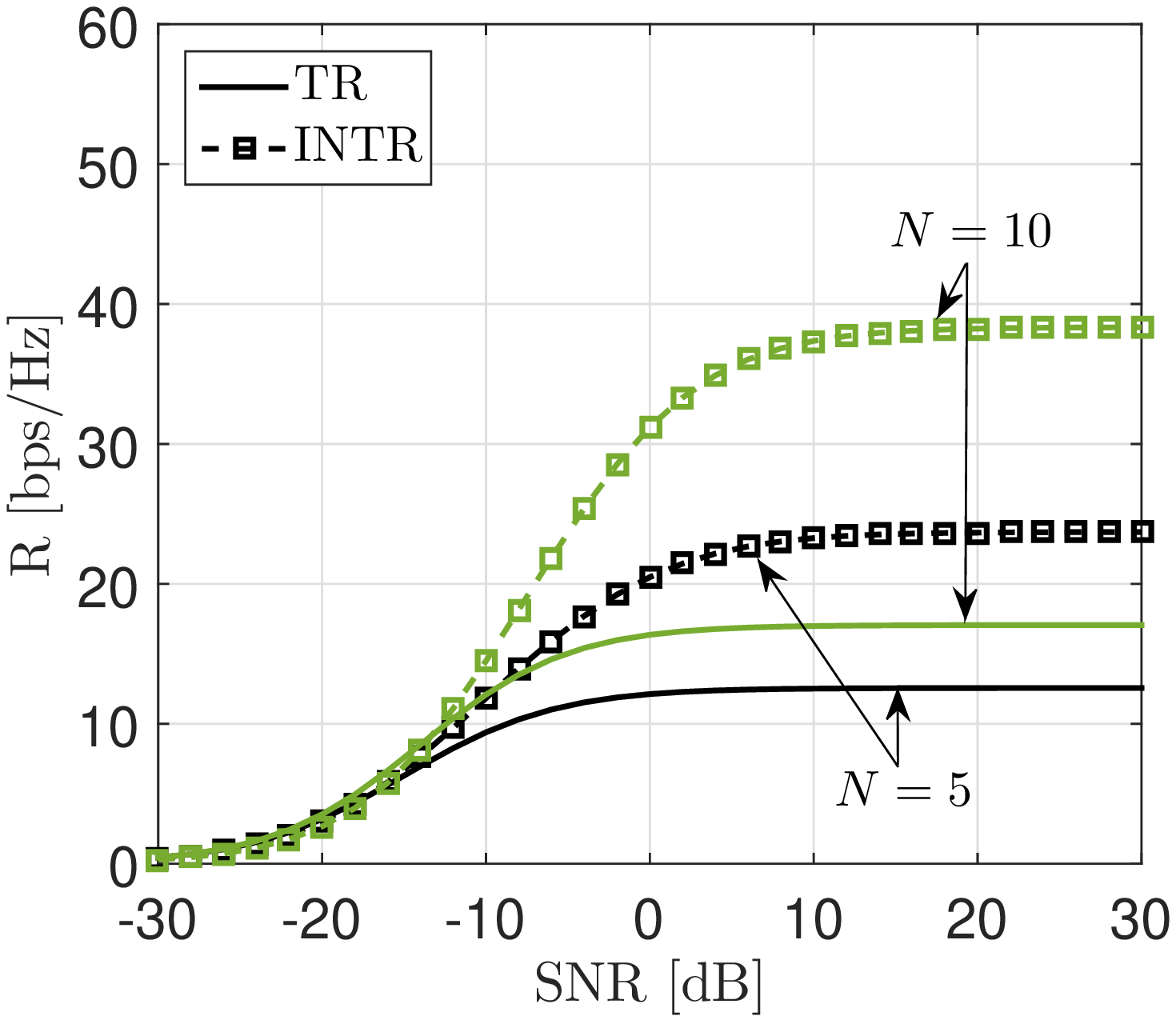}
\label{fig_ratem32}}
\subfloat[]{\includegraphics[width=0.33\columnwidth]{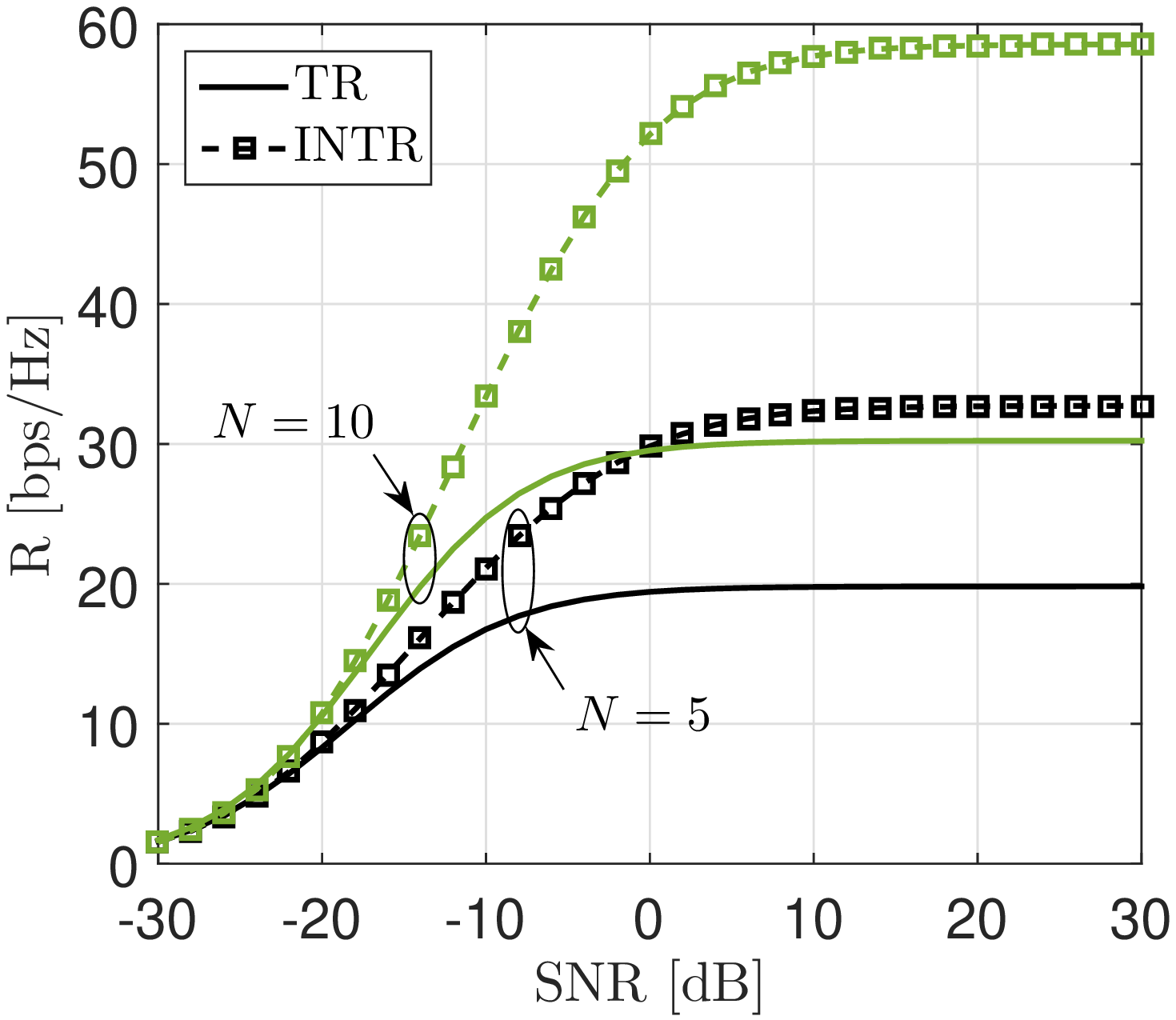}
\label{fig_ratem128}}
\subfloat[]{\includegraphics[width=0.33\columnwidth]{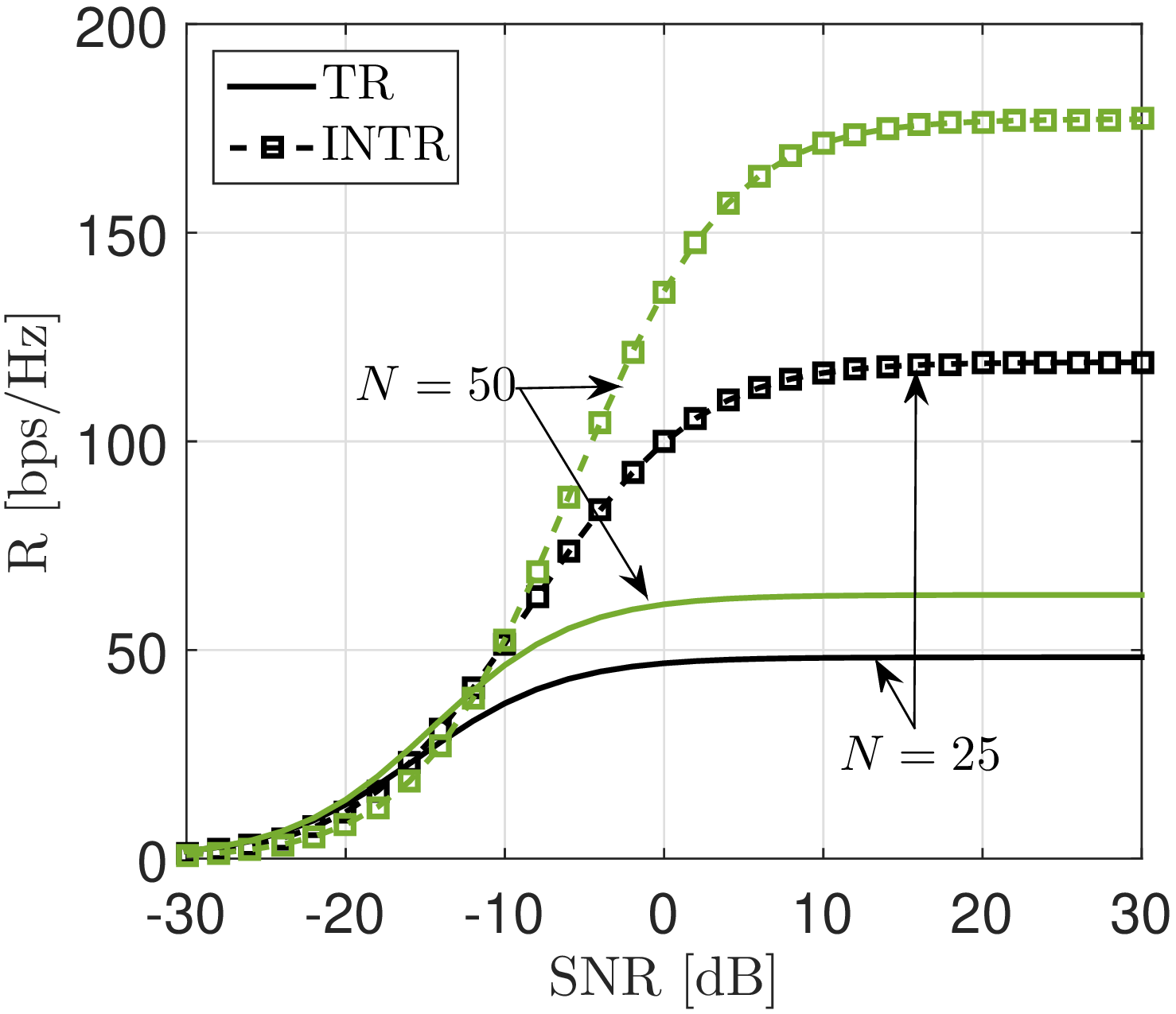}
\label{fig_ratem128x}}
\newline
\caption{Achievable rate of TR and INTR in the LR scenario. (a) $M=32$ antennas, (b) $M=128$ antennas, (c) $M=128$ antennas with an extreme number of users. The multiplexing gain increases with the number of antennas.}
\label{fig_rate}
\end{figure*}

The achievable sum rate measures the downlink spectral efficiency in a multiple-access system. Assuming that each user treats ISI and IUI as Gaussian interferences, and according to the model defined in Sections \ref{Model} and \ref{Model2}, the average achievable sum rate for a multi-user TR system is
\begin{IEEEeqnarray}{rCl}
\label{eq_sumrate}
R & = & \mathbb{E} \left[ \sum_{n=1}^N \log_2{\left( 1+\frac{P_{s,n}}{P_{isi,n}+P_{iui,n}+\sigma_z^2} \right)} \right],
\end{IEEEeqnarray}
where $P_{s,n}$, $P_{isi,n}$, and $P_{iui,n}$ are the desired signal power, ISI power, and IUI power, respectively, calculated at user $n$ for a given realization. For simplicity, it is also assumed that the channel is used for downlink transmission all the time. A proper reduction factor can be used to account for uplink time in a TDD system or channel estimation overheads. Numerical results for the average achievable sum rate are shown in Fig. \ref{fig_rate}. These results were obtained in the LR scenario with correlated channels and $L_p = 90$. As seen, INTR offers a significant improvement over conventional TR, doubling its rate in some cases and providing a remarkable multiplexing gain. In addition, we simulated a more extreme case with $N=30$, and $N=50$ and 128 antennas, with the purpose of further demonstrate the capabilities of TR to handle IUI. Results are shown Fig. \ref{fig_ratem128x}. Even though the assumption of uncorrelated CIR between users is hardly met when $N$ is that large, results show that an outstanding efficiency of more than 170 bps/Hz can be achieved with INTR. In all the simulated scenarios our proposed INTR technique outperforms conventional TR, as it can better withstand an increase in user load.

\section{Conclusion}
\label{conclusion}

We have analyzed a baseband TR beamforming system for mm-wave multi-user massive MIMO. We studied conventional TR and equalized TR and found that their performance is IUI limited. We also noticed that, when the number of antennas is large, the ratio between the desired signal power and ISI or IUI power increases. Thus, we confirm the potential of TR as a beamforming technology for massive MIMO. We also note that equalizing solutions such as ETR are not necessary when the number of transmit antennas is large. After identifying IUI as the main detection impairment for TR systems, we propose a modified technique called INTR. This technique calculates the transmit pre-filters in the frequency domain that set the IUI to zero and are closest to the original TR solution. We proposed a 60 GHz MIMO channel model, where CIR taps are modeled with Nakagami distributed amplitudes. In addition, we use PDPs given by the IEEE 802.11ad SISO NLoS model, and generate spatial correlation in the CIRs according to a geometrical model. By means of numerical simulations, we verified that the proposed INTR outperforms conventional TR with respect to average BER and achievable sum rate. In particular, we note that INTR performance is extremely tolerant to increases in the number of users, and provides both diversity and multiplexing gains simultaneously.

\bibliographystyle{IEEEtran}
\bibliography{MUTR}

\end{document}